# Title: Biogenic magnetic nanoparticles in plants

**Authors**: S.V. Gorobets[1], O. Yu. Gorobets[1,2*], A.V. Magerman[1], Yu. I. Gorobets[2], I.V. Sharay[2]

**Affiliations:**

[1]National Technical University of Ukraine «Igor Sikorsky Kyiv Polytechnic Institute», 37 Peremohy Ave., 03056 Kyiv, Ukraine.

[2]Institute of Magnetism of NAS and MES of Ukraine, 36b Acad. Vernadskoho Blvd., 03142 Kyiv, Ukraine.

*Correspondence to: Email: gorobets.oksana@gmail.com.

**Abstract:** The genetic programming of biosynthesis of biogenic magnetic nanoparticles (BMNs) in plants was revealed by methods of comparative genomics. The samples of leaves and the root of *Nicotiana tabacum*, the stems and tubers of *Solanum tuberosum* and the stems of pea *Pisum sativum* were examined by scanning probe microscopy (in atomic force and magnetic power modes), and it was found that the BMNs are located in the form of chains in the wall of the phloem sieve tubes (ie, the vascular tissue of plants). Such a localization of BMNs supports the idea that the chains of BMNs in different organs of plants have common metabolic functions. Stray gradient magnetic fields about several thousand Oe, which are created by chains of BMNs, can significantly affect the processes of mass transfer near the membrane of vesicles, granules, organelles, structural elements of the membrane, and others. This process is enhanced in plants when artificial magnetite is added to the soil.

**One Sentence Summary:** Biogenic magnetic nanoparticles are revealed in plants.

**Main Text:** It is known that the magnetic fields that provide the electrical processes in organisms are extremely small even compared with the magnitude of the magnetic field of the Earth, and therefore the biomagnetic phenomena associated with the influence of their own magnetic fields in living organisms on their metabolism, in contrast to bioelectric phenomena, are practically not investigated (1-3). This point of view had been persisting for more than five decades, even after the discovery by Lowenstam of strong natural nanoscale magnets (biogenic magnetic nanoparticles (BMNs)) in teeth of mollusks (4) in 1962, and the discovery by Blackmore of such bacteria intracellular particles (5) in 1975, the synthesis of which is genetically programmed and carried out by microorganisms by themselves. To date, BMNs has been experimentally detected in algae and protozoa (6), worms (7), coats (8), snails (9), ant and butterflies (10-12), honey bees (13), termites (14), lobsters (15), tritons (16), migratory and non-migratory fish (17-21), turtles (22, 23), birds (24-27), bats (28), dolphins and whales (29), pigs (30) and humans (31-36). In humans, BMNs is detected both in norm and in pathologies, for example, BMNs are found in neurodegenerative diseases (37-39), oncological diseases (31, 40), heart aneurysms (41), atherosclerosis (42). The methods of comparative genomics have shown that the genetic apparatus of the BMNs biosynthesis is unique in the representatives of all kingdoms of living organisms and is based on genes that originate from a common ancestor before the appearance of multicellular organisms (43-46).

However, due to the fact that for many years BMNs was investigated mainly in connection with the ideas about magnetotaxis and magnetoreception, BMNs, which are the source of their own gradient magnetic fields, were almost not studied in plants. In this case, the gradients of magnetostatic stray fields of the BMN have a sufficient value ($\frac{\nabla \vec{H}^2}{2} \Box (10^{10} \div 10^{12}) \frac{Oe^2}{cm}$) to influence the transport of cells and their components, vesicles, granules, etc (1, 43,44, 47). Moreover, the displacement of intracellular amyloplast was observed experimentally under the influence of an external gradient magnetic field, even with a significantly lower dynamic factor $\frac{\nabla \vec{H}^2}{2} \Box (10^{9} \div 10^{10}) \frac{Oe^2}{cm}$, and, as a consequence of the distortion of the seedlings of the barley Hordeum vulgare in the direction of the gradient of the magnetic field (48). In this regard, the purpose of this work is to identify plants that are potential producers of BMNs by methods of comparative genomics, to study experimentally the presence of BMNs in the samples of plants-potential producers of BMNs and to study the effects of artificial magnetite on the development of plants.

Results and discussion: Comparison of amino acid sequences of the proteins of the Mam group, without which the biomineralization of the BMNs in *Magnetospirillum gryphiswaldense* MSR1 (49) is not possible with genomes of plants, deciphered 50% or more in the GenBank NCBI (50) is carried out. The following genera of angiosperms (40 plants) were investigated: Fabaceae, Brassicaceae, Cucurbitaceae, Rosaceae, Rutaceae, Solanaceae, Amaranthaceae, Apiaceae, Poaceae, Bromeliaceae, Arecaceae, Chenopodioideae, Pedaliaceae, Salicaceae, Betulaceae, Vitaceae, Linaceae, Oleaceae, Malvaceae. The following plant divisions (15 plants) were also investigated: Lycopodiophyta, Chlorophyta, Phaeophyceae and Bacillariophyceae. For example, typical alignment results for a number of investigated plants are presented in Table.1. Investigated angiosperms are widespread crop plants used in agriculture.

Table 1. Comparison of Mam group proteins of the MTB *Magnetospirillum gryphiswaldense* MSR-1 with proteomes of plants.

| Plant | Comp-lexity of the genome | | *Magnetospirillum gryphiswaldense* MSR-1 | | | | | |
|---|---|---|---|---|---|---|---|---|
| | | | MamA | MamB | MamM | MamO | MamE | MamK |
| Angiosperms | | | | | | | | |
| *Arabidopsis thaliana* | ● | E-value | 9·10⁻⁶ | 6·10⁻³⁵ | 7·10⁻²³ | 3·10⁻⁷ | 8·10⁻³³ | 0.004 |
| | | Ident% | 26 | 28 | 28 | 25 | 45 | 22 |
| | | Length | 169 | 303 | 273 | 175 | 162 | 199 |
| *Nicotiana tabacum* (tobacco) | ◐ | E-value | 3·10⁻⁵ | 3·10⁻²⁹ | 6·10⁻²⁶ | 1·10⁻⁷ | 2·10⁻³³ | 0.003 |
| | | Ident,% | 28 | 27 | 30 | 25 | 46 | 22 |
| | | Length | 137 | 307 | 301 | 175 | 162 | 197 |
| *Solanum tuberosum* (potato) | ● | E-value | 2·10⁻⁵ | 2·10⁻³⁰ | 2·10⁻²⁶ | 7·10⁻⁷ | 4·10⁻³² | 2·10⁻⁴ |
| | | Ident,% | 23 | 27 | 30 | 25 | 45 | 21 |
| | | Length | 130 | 313 | 297 | 175 | 162 | 215 |
| *Pisum sativum* (pea) | ● | E-value | 7·10⁻⁵ | 3·10⁻³¹ | 2·10⁻²⁸ | 3·10⁻⁶ | 7·10⁻³² | 0.006 |
| | | Ident,% | 33 | 27 | 30 | 24 | 44 | 22 |
| | | Length | 76 | 293 | 281 | 185 | 162 | 174 |
| | | Length | 169 | 296 | 290 | 193 | 162 | 199 |

| Lycopodiophyta | | | | | | | | |
|---|---|---|---|---|---|---|---|---|
| *Selaginella kraussiana* | | E-value | $2 \cdot 10^{-5}$ | $6 \cdot 10^{-36}$ | $4 \cdot 10^{-32}$ | $7 \cdot 10^{-8}$ | $9 \cdot 10^{-35}$ | 0.005 |
| | | Ident | 26 | 33 | 31 | 23 | 43 | 25 |
| | | Length | 933 | 327 | 327 | 444 | 413 | 152 |
| Chlorophyta | | | | | | | | |
| *Phormidium tenue* NIES-30 | | E-value | $2 \cdot 10^{-10}$ | $3 \cdot 10^{-18}$ | $9 \cdot 10^{-25}$ | $1 \cdot 10^{-6}$ | $4 \cdot 10^{-33}$ | $4 \cdot 10^{-12}$ |
| | | Ident | 28 | 22 | 28 | 28 | 46 | 26 |
| | | Length | 133 | 280 | 260 | 148 | 168 | 310 |
| Phaeophyceae | | | | | | | | |
| *Ectocarpus siliculosus (taxid:28)* | | E-value | $6 \cdot 10^{-7}$ | $5 \cdot 10^{-39}$ | $1 \cdot 10^{-27}$ | $5 \cdot 10^{-6}$ | $3 \cdot 10^{-37}$ | 0.013 |
| | | Ident | 35 | 28 | 30 | 24 | 43 | 29 |
| | | Length | 69 | 307 | 271 | 165 | 167 | 69 |
| Bacillariophyceae | | | | | | | | |
| *Thalassiosira pseudonana* | | E-value | $2 \cdot 10^{-9}$ | $9 \cdot 10^{-32}$ | $2 \cdot 10^{-17}$ | $5 \cdot 10^{-9}$ | $1 \cdot 10^{-34}$ | 0.004 |
| | | Ident | 25 | 29 | 26 | 22 | 41 | 23 |
| | | Length | 175 | 261 | 288 | 181 | 171 | 180 |

The bioinformatic analysis showed that all investigated 55 plants, which were examined, are potential producers of BMNs. Methods of pairwise alignment of amino acid sequences were used for the study by using the "BLAST" NCBI that is in free access. The following standard parameters were used for the analysis of the results of research (51): the value of the E-number (that is, the number of alignments with such or better alignment weight that can be found by chance in a database of a certain size), Ident – the percentage of overlapping of amino acid sequences within which the alignment is made, Length – the number of identical amino acid residues of the compared proteins, with optimal alignment and the function of the aligned proteins. The value of the statistical numbers (E-number, Ident, Length) that were used to evaluate the protein homology and the comparison of the functions of the biomineralization proteins are in the same range of values in plants and the MTB *Magnetospirillum gryphiswaldense* MSR-1 as in the alignment obtained for biomineralization proteins of human and animal (44,46, 52), non-magnetitoxic bacteria (53-59) and proteins of MTB *Magnetospirillum gryphiswaldense* MSR-1. In addition, the functions of Mam proteins (the biomineralization proteins of the BMNs of MTB *Magnetospirillum gryphiswaldense* MSR-1) and the functions of the homologous proteins in the plants coincide (Table 2). This confirms the hypothesis of the origin of the biomineralization proteins from a common ancestor at the dawn of evolution (44-46).

Table 2. Functions of the Mam proteins of MTB *Magnetospirillum gryphiswaldense* MSR-1 and the functions of the homologous proteins in the plants.

| Protein of MTB | Functions of a Mam protein | Name and functions of the homologous proteins in the plants |
|---|---|---|
| MamA | Contains the TPR domain, which is a consensus sequence. The TPR domain is involved in a variety of functions, including protein-protein interactions, chaperone | Pex5-protein – peroxisome – is widespread, surrounded by an organoid cell membrane with a large variety of metabolic functions - the destruction of toxic compounds, the construction of myelin |

|  | functions, cell cycle, transcription, transport of proteins | sheath of nerve fibers, etc. Enzymes of organelles use molecular oxygen to split hydrogen atoms from non-organic substrates to form peroxide. Pex5 contains the TPR domain. |
|---|---|---|
| MamB, MamM | Transport of cations Co, Zn, Cd, Fe, Ni. | (CDF) Cation efflux family protein – integral membrane proteins that increase tolerance to ion bivalent metals such as cadmium, zinc and cobalt et al. |
| MamE, MamO | Serine protease. The PDZ domain of the trypsin-like serine protease is involved in the response to heat shock, the function of chaperones, apoptosis, may be responsible for recognizing the substrate and / or binding. | DegP protease 1 – serine protease. Serine proteases possess a wide spectrum of peptidase activity, including exopeptidase, endopeptidase, oligopeptidase and omega peptidase activity. Serine proteases are involved in important physiological processes, including regulation of mitochondrial homeostasis, apoptosis, and transmission of cellular signals involved in the development of pathological processes. |
|  |  |  |

Methods of atomic force microscopy and magnetic force microscopy were used for experimental study of plant tissues for the presence of them in BMNs. The study of BMNs in plants was carried out on the samples of tobacco *Nicotiana tabacum*, as the most studied model organism among plants, as well as pea *Pisum sativum* and potato *Solanum tuberosum*.

Tobacco grown in accordance with the methodology (60) on the nutrient medium Murasige-Scuga. Such a choice of nutrient medium is due to the fact that it does not contain magnetite nanoparticles, in contrast to the vast majority of soils, which usually contain a concentration of magnetite nanoparticles of 2-6% by weight of soil (61).

A leaf (Fig. 1, a1-a3), a leaf vein (Fig. 1, b1-b3) and a root (Fig. 1, c1-c3) were investigated in the tobacco to check the presence of BMNs. BMNs are located on the membrane of sieve tubes of phloem in a leaf, a leaf vein and a roots of tobacco (Fig. 1). The phloem is a vascular tissue of plants that forms a network of sieve tubes through which the transport of organic substances synthesized by leaves during photosynthesis is provided to all organs of the plant (62), in contrast to the vascular tissue of plants – xylem, which provides transport of water and mineral substances from the soil (63). The sieve tubes of the tobacco leaf, that are shown in Figure 1, have typical morphology and dimensions as described in (64).

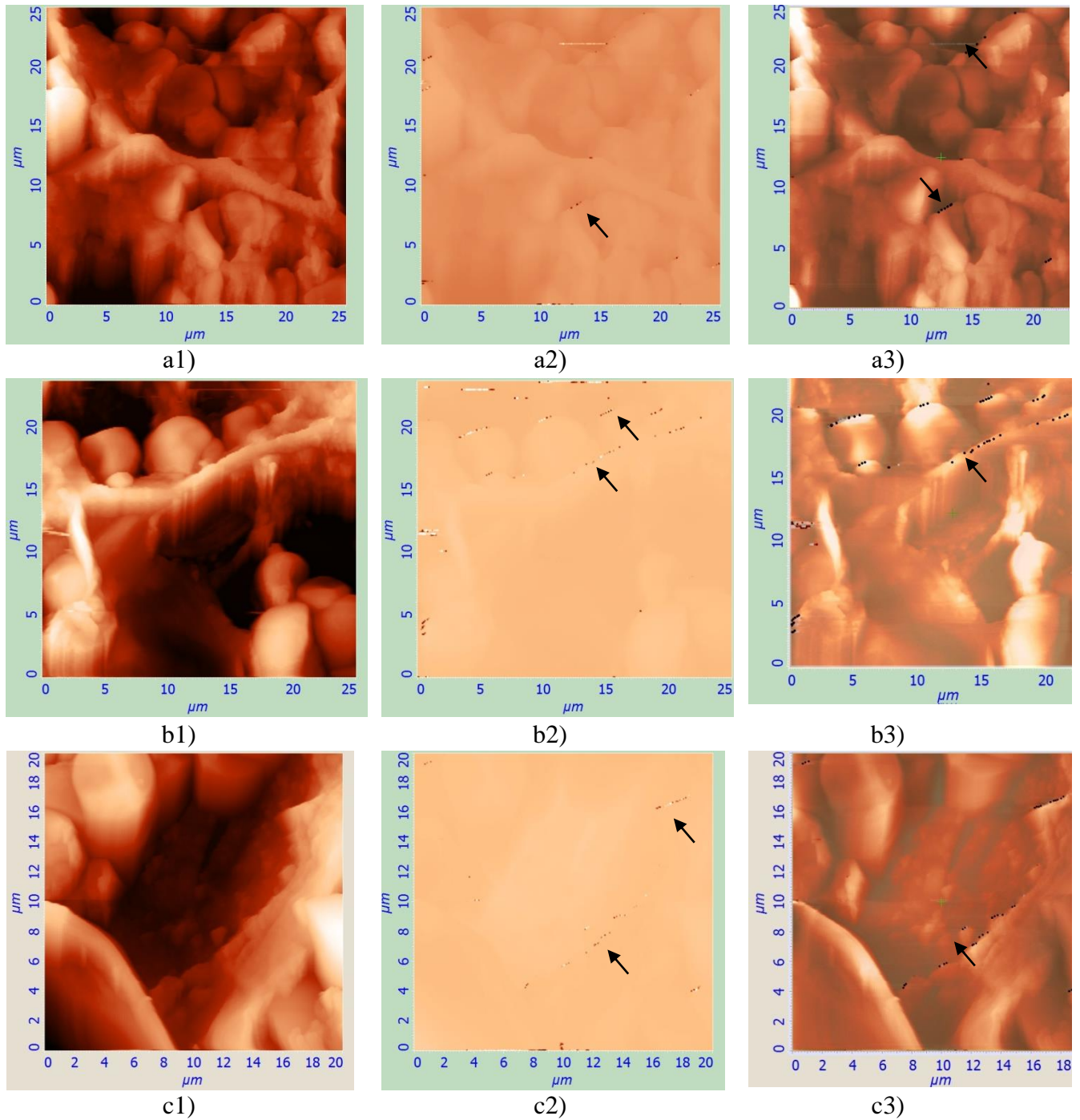

Fig. 1. Scanning probe microscopy of tobacco *Nicotiana tabacum*: a1) – AFM image of tobacco leaf , a2) – MFM image of tobacco leaf (BMNs are shown with arrows), a3) – combined AFM and MFM images of tobacco leaf (arrow indicate pores of sieve tubes); b1) – AFM image of a vein of tobacco leaf , b2) – MFM image of a vein of tobacco leaf (BMNs are shown with arrows), b3) – combined AFM and MFM images of a vein of tobacco leaf (arrow indicate membranes of sieve tubes); c1) – AFM image of a root of tobacco, c2) – MFM image of a root of tobacco (BMNs are shown with arrows), c3) – combined AFM and MFM images of a root of tobacco (arrows indicate sieve tube).

A similar localization of BMNs is observed in samples of potato (Fig. 2). It can be seen from Fig. 2 that BMNS in the stem (Fig. 2, a1-a3) and potato tubers (Fig. 2, b1-b3) are associated with short chains located on the boundary of the vascular tissue in the potato stem (Fig. 2, a3) and along the boundaries of starch grains and sieve tubes (Fig. 2, b3) in potato tubers (65).

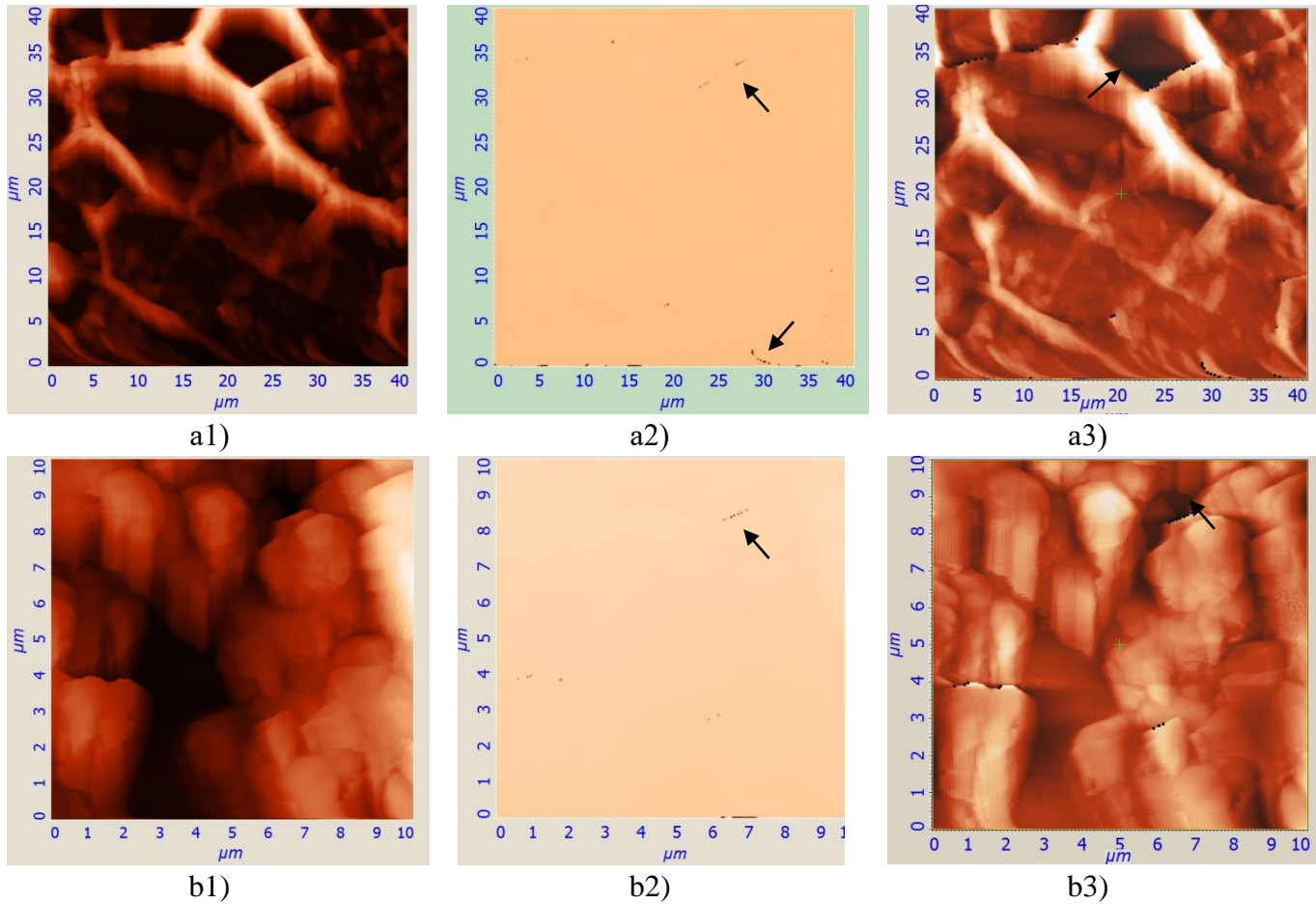

Fig. 2. Scanning probe microscopy of potato *Solanum tuberosum*: a1) – AFM images of a stem of potato; a2) – MFM images of a stem of potato (BMNs are shown with arrows), a3) – combined AFM and MFM images of a stem of potato (arrows indicate pores of sieve tubes); b1) – AFM image of a tubers of potato, b2) – MFM image of tubers of potato (BMNs are shown with arrows), b3) – combined AFM and MFM images of potato tubers (arrows indicate pores of sieve tubes).

The location of BMNs in plants at the boundary of the vascular tissue is similar to the location of BMNs, in the samples of fungi on the cell walls of hyphae (66), in the tissues and organs of animals (including humans) on the walls of the capillaries (41) from a functional point of view.

The features of the growth of pea *Pisum sativum*, grown on soils without the addition of magnetite and on soils with the addition of magnetite nanoparticles in a magnetic fluid (with an average size of nanoparticles of 11 nm, a minimum particle size of 2 nm and a maximum size of 23 nm) were studied. The magnetic fluid is obtained by the method (67). Magnetite was used at concentrations of 1 mg / ml and 0.1 mg / ml (concentration close to the content of magnetite in the soil (61)).

BMNs and nanoparticles of artificial magnetite in pea *Pisum sativum* (Fig. 3) are located on the membrane of phloem sieve tubes, as well as in the samples of tobacco *Nicotiana tabacum* (Fig.1) and potato *Solanum tuberosum* (Fig. 2).

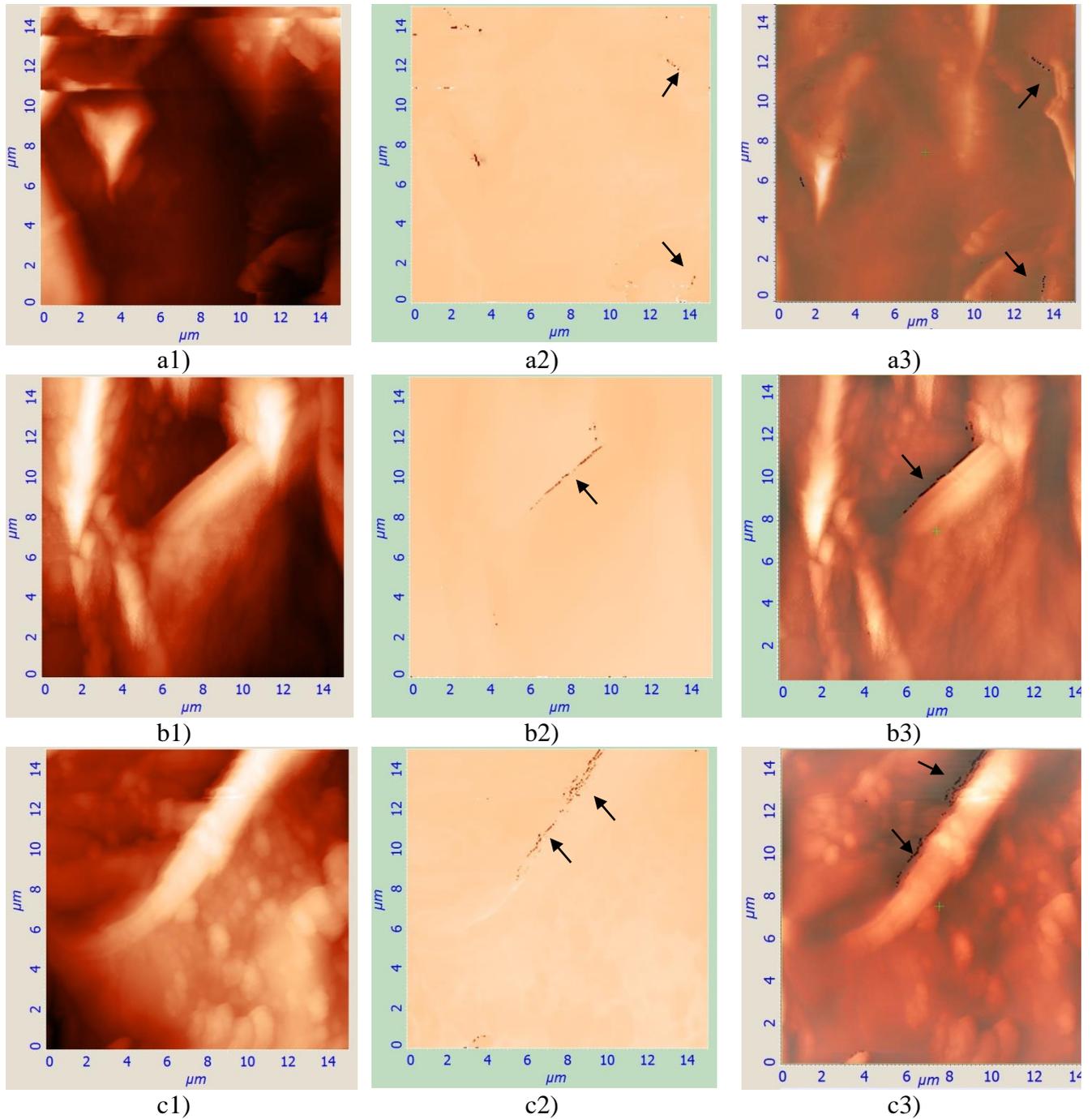

Fig. 3. Scanning probe microscopy of the stem of pea *Pisum sativum*: a1) – AFM images of a stem of pea grown on the soil without the addition of magnetite; a2) – MFM images of a stem of pea grown on the soil without the addition of magnetite (BMNs and nanoparticles of artificial magnetite are shown with arrows), a3) – combined AFM and MFM images of pea stem grown on the soil without the addition of magnetite (arrows indicate pores of sieve tubes); b1) – AFM image of a stem of pea grown on the soil with the addition of magnetite (concentration of 0.1 mg/ml), b2) – MFM image of a stem of pea grown on the soil with the addition of magnetite (concentration of 0.1 mg/ml) (BMNs and nanoparticles of artificial magnetite are shown with arrow), b3) – combined AFM and MFM images of a stem of pea grown on the soil with the addition of magnetite (concentration of 0.1 mg/ml) (arrow indicate sieve tube); c1) – AFM image of a stem of pea grown on the soil with the addition of magnetite (concentration of 1 mg/ml), c2) – MFM image of a stem of pea grown on the soil with the addition of magnetite (concentration of 1 mg/ml) (BMNs and nanoparticles of artificial magnetite are shown with arrows), c3) – combined AFM and MFM images of a stem of pea grown on the soil with the addition of magnetite (concentration of 1 mg/ml) (arrows indicate pores of sieve tubes).

Based on the results of AFM and MFM, the maximum size of the BMNs (as the average distance between adjacent black or white areas in Figure 1, Figure 2) and the amount of BMNs were estimated in the chain of examined tissues of tobacco, potato and pea (Table 3).

Comparing the results of AFM and MFM of known producers of BMNs, the maximum size of BMNs and the amount of BMNs in the chain of investigated organisms were estimated for comparison with the relevant data for the model organism, namely the magnetotaxis bacterium *Magnetospirillum gryphiswaldense* MSR-1 (Table 3).

Table 3. Sizes of BMNs in tobacco and potato and comparison with literature data for other organisms.

| Organism | Estimation of the maximum size of BMNs, nm | Number of particles in chains |
|---|---|---|
| Tobacco leaf *Nicotiana tabacum* | 110-220 | 4-10 |
| Tobacco root *Nicotiana tabacum* | 80-185 | 6-10 |
| Potato stem *Solanum tuberosum* | 60-120 | 4-8 |
| Potato tuber *Solanum tuberosum* | 35-60 | 2-8 |
| Pea stem *Pisum sativum* (control) | 95-105 | 3-7 |
| Mushroom *Agaricus bisporus* | 55-85 (66) | 2-7 (66) |
| Magnetotaxis bacteria *Magnetospirillum gryphiswaldense* MSR-1 | 10-40, 35-120 (68) | 4-200 (68) |
| Termite | ≈10 (69) | |
| Beak of *Gallus gallus domesticus*, *Columba livia*, *Erithacus rubecula* | ~1000 (70,71) | 10-15 (70,71) |
| The brain of carp *Cyprinus carpio* | 350-405 (66) | ≈12 (66) |
| Human cerebral cortex | 90-200 (72) | ≈80 (72) |

When growing pea *Pisum sativum* with the addition of artificial nanoparticles of magnetite in the soil, on the walls of sieve tubes conglomerates of nanoparticles, which include both BMNs and artificial nanoparticles of magnetite, are formed (Figs 3b,3c). In this case, the number and size of the formed magnetite conglomerates and the number of their chains (Fig. 3c, 3c) are different from the control (Fig. 3a, Table 4).

Table 4. The number and size of BMNs and artificial nanoparticles of magnetite in the stem of pea *Pisum sativum,* grown with the addition of artificial magnetite nanoparticles in the soil.

| Conditions for growing pea *Pisum sativum* | Estimation of the maximum size of the BMNs, nm | Number of BMNs and artificial magnetite nanoparticles in chains, pcs | Number of chains of BMNs and artificial magnetite nanoparticles parallel to chains of BMNs, pcs |
|---|---|---|---|
| Soil without the addition of magnetite | 95-105 | 3-7 | 0 |
| Soil with the addition of magnetite (concentration 0.1 mg/ml) | 95-100 | 7-13 | 0 |
| Soil with the addition of magnetite (concentration 1 mg/ml) | 94-100 | 6-14 | 1-2 |

From Fig. 3 and Table 4 it is evident that when the concentration of magnetite in the soil increases, the amount of magnetic nanoparticles in the chain increases, which proves that an artificial magnetite can be embedded in a chain of biogenic magnetic nanoparticles or form additional chains (Figs 3c, 3c). We observe not only complexes of biogenic and artificial magnetite (Fig. 3b), located on membranes of sieve tubes and also parallel chains of artificial nanoparticles (Fig. 3c) at concentration of magnetite 1 mg/ml. The formation of chains of artificial magnetite nanoparticles parallel to the BMNs chain leads to a change in the spatial distribution of magnetostatic fields in the vicinity of natural BMNs. With significant accumulation of the chains of artificial nanoparticles of magnetite, they can serve as a magnetic core and causes the magnetic field lines of BMNs to be concentrated in the core material, which is correlated with the results of plant growth (Table 5, Fig. 4).

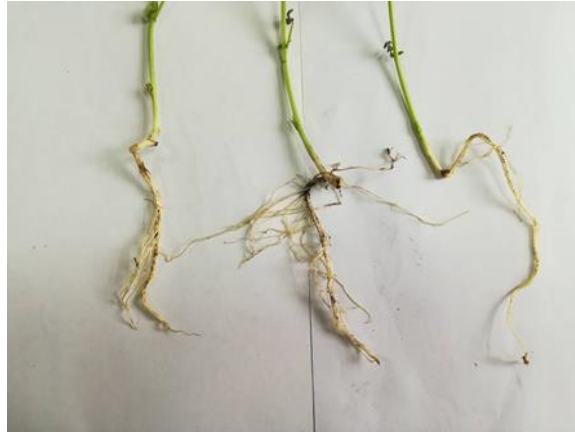

Figure 4. Morphology of roots of pea *Pisum sativum* (from left to right): the plant grown on the soil without the addition of magnetite (control), the plant grown on the soil with the addition of magnetite (concentration of 0.1 mg/ml), the plant grown on the soil with the addition of magnetite (concentration of 1 mg/ml).

Table 5. Influence of different concentrations of magnetite on the morphology of *pea Pisum sativum*.

| Average values | Control | 0.1 mg/ml | 1 mg/ml |
|---|---|---|---|
| Length of stems, cm | 25,2±1,2 | 33,8±0,8 (*34%) | 26,8±0,9 (*6%) |
| Length of roots, cm | 9,4±0,2 | 8,6±0,3 (*-9%) | 13,1±1,1 (*39%) |
| Length of plants, cm | 35±2 | 42,3±1,6 (*22%) | 39,8±1,9 (*15%) |
| Length of leaves, cm | 1,3±0,3 | 1,9±0,4 (*46%) | 1,1±0,3 (*-16%) |
| Number of lateral roots, pc | 8±1 | 31±2 (*287%) | 4±2 (*-50%) |
| Number of leaves, pc | 24±2 | 29±1 (*20%) | 29±2 (*20%) |
| Weight of plants, g | 8,64 | 11,54 | 9,68 |
| Weight of roots, g | 1,75 | 2,3 | 1.7 |

* increase/decrease in length, weight of plants and the number of roots and leaves in relation to the control

Morphological differences of plants of pea *Pisum sativum*, grown on the soil with the addition of artificial nanoparticles of magnetite at a concentration of 0.1 mg/ml (Fig. 4, Table 5), are similar to morphological changes occurring in plants with an increase in the intensity of the synthesis of phytohormones auxin, which affects the formation and growth of roots (73) namely there is a shortening of the main root and the development of lateral roots in both cases. It is known that phytohormone auxin is transported by a plant with vesicle sizes 180-220 nm (74).

Thus, it is likely that with considerable accumulation of artificial nanoparticles of magnetite and closure of magnetic fields, gradient magnetic forces in the vicinity of the BMNs, acting on vesicles, granules and liposomes, will be significantly reduced, which affects the growth of plants (Fig. 4, Table 5).

Based on experimental data and theoretical calculations (44,47,75), it can be supposed that gradient magnetic forces in the vicinity of the BMNs are sufficient for the accumulation of vesicles, granules, liposomes (47, 75), amiloplasts (76) to hold the vesicles in the vicinity of the BMNs chain for BMNs sizes from 20 nm to 150 nm and the size of vesicles greater than

100 nm, that is, near the membrane (44,47). Since the size of vesicles and BMNs in plants are in this range (Table 3), it can be shown that the BMNs perform the same function in the plant organism as in human and animals, namely the function of concentrators of vesicles, granules and other biological objects, including vesicular transport.

**Conclusion:** It has been shown by the methods of comparative genomics that all species are potential producers of intracellular crystalline BMNs among the investigated plants with genomes, deciphered by more than 50%.

In this case, experimental studies of BMNs in plant samples, carried out in this work by AFM and MFM methods, showed that: BMNs in plants form chains and BMNs in plant organisms are the part of the transport system. So, the BMNs in the plants are located in the wall of the vascular tissue, namely in the wall of the sieve tubes of the phloem.

Chains of BMNs are the components of cells that form the walls of vascular tissue – in the wall of sieve tubes phloem. In this case, the vascular tissue of phloem of plants serves for the transfer of organic substances, hormones, etc. (62). The same localization of BMNs chains (namely, in the wall of the vascular tissue of phloem) in different organs of higher plants cannot be occasional, taking into account that the genetically engineered biosynthesis mechanism of BMNs appeared at the beginning of evolution (77). Such a localization of BMNs suggests the benefit of the idea that the chains of BMS have common metabolic functions in different organs of plants. It has already been mentioned that the BMNs chains create stray magnetic fields of several thousand Oe and magnetic field gradients that can significantly affect the processes of mass transfer near the membrane of vesicles, (47,44), it can be supposed that gradient magnetic forces in the vicinity of) organelles, structural elements of the membrane and others. With significant accumulation of chains of artificial nanoparticles of magnetite in the vicinity of BMNs, further formed chains of magnetite can serve as a magnetic circuit and causes the scattering of magnetic field lines of BMNs to be concentrated in the core material, which can lead to a weakening or exclusion of the function of BMNs as a magnetic nano-device.